\documentclass[preprintnumbers]{article}
\usepackage[utf8]{inputenc}
\usepackage{amsmath,graphicx}
\usepackage[numbers,sort&compress]{natbib}
\usepackage{hyperref}

\begin{document}
\rightline{FERMILAB-FN-1157, LA-UR-22-22524}
\bigskip

\centerline{\bf \large SBN-BD: $\mathcal{O}$(10 GeV) Proton Beam Dump at Fermilab’s PIP-II Linac}

\bigskip
\centerline{\bf \large Contacts}
\centerline{Matt Toups (FNAL) [toups@fnal.gov], R.G. Van de Water (LANL) [vdwater@lanl.gov]}
\medskip

\centerline{\bf \large Authors and Proponents}
\centerline{Brian Batell (University of Pittsburgh), S.J. Brice (FNAL), Patrick deNiverville (LANL), }
\centerline{Jeff Eldred (FNAL),  A. Fava (FNAL), Kevin J. Kelly (FNAL), Tom Kobilarcik (FNAL),}
\centerline{W.C. Louis (LANL),   Pedro A.~N. Machado (FNAL), Z. Pavlovic (FNAL), Bill Pellico (FNAL), }
\centerline{Josh Spitz (University of Michigan), Rex Tayloe (Indiana University), R. T. Thornton (LANL), }
\centerline{Jaehoon Yu (University of Texas at Arlington), J. Zettlemoyer (FNAL)}

\bigskip
\centerline{\large March, 2022 }

\begin{abstract}
  Proton beam dumps are prolific sources of mesons enabling a powerful technique to search for vector mediator coupling of dark matter to neutral pion and higher mass meson decays.  By the end of the decade the PIP-II linac will be delivering up to 1 MW of proton power to the FNAL campus.  This includes a significant increase of power to the Booster Neutrino Beamline (BNB) which delivers 8 GeV protons to the Short Baseline Neutrino (SBN) detectors.  By building a new dedicated beam dump target station, and using the SBN detectors, a greater than an order of  magnitude increase in search sensitivity for dark matter relative to the recent MiniBooNE beam dump search can be achieved.  This modest cost upgrade to the BNB would begin testing models of the highly motivated relic density limit predictions and provide novel ways to test explanations of the anomalous excess of low energy events seen by MiniBooNE.

\end{abstract}

\section{Physics Goals and Motivation}

Recent theoretical work has highlighted the motivations for
sub-GeV dark matter candidates that interact with ordinary matter
through new light mediator particles \cite{DS,CVision,BRN}.
These scenarios constitute a cosmologically and 
phenomenologically viable possibility to account for the dark matter 
of the universe. Such sub-GeV (or light) dark matter particles are difficult to 
probe using traditional methods of dark matter detection, but can be 
copiously produced and then detected with neutrino beam experiments 
such as MiniBooNE, the Short Baseline Neutrino (SBN) Program, NOvA, and DUNE \cite{NuDM}.  This represents a new 
experimental approach to search for dark matter and  
is highly complementary to other approaches such as underground direct 
detection experiments, cosmic and gamma ray satellite and balloon 
experiments, neutrino telescopes, and high energy collider 
experiments~\cite{DS,CVision,BRN}.  Furthermore, searches for light dark matter provide an 
additional important physics motivation for the current and future 
experimental particle physics research program at the Fermi National Accelerator Laboratory (FNAL).  

The MiniBooNE experiment running at the  FNAL Booster Neutrino Beamline (BNB) was originally designed for neutrino
oscillation and cross section measurements.  In 2014 a special beam dump run was carried out which suppressed neutrino produced backgrounds while enhancing the search for sub-GeV dark matter via neutral current scattering, resulting in new significant sub-GeV dark matter limits \cite{MiniBooNE:2017nqe,MB-DM}.  The result clearly demonstrated the unique and powerful ability to search for dark matter with a beam dump neutrino experiment.  

\section{A New BNB Beam Dump Target Station and Running in the PIP-II Era}

\begin{figure}[ht]
 \centerline{ 
 \includegraphics[width=0.45\textwidth]{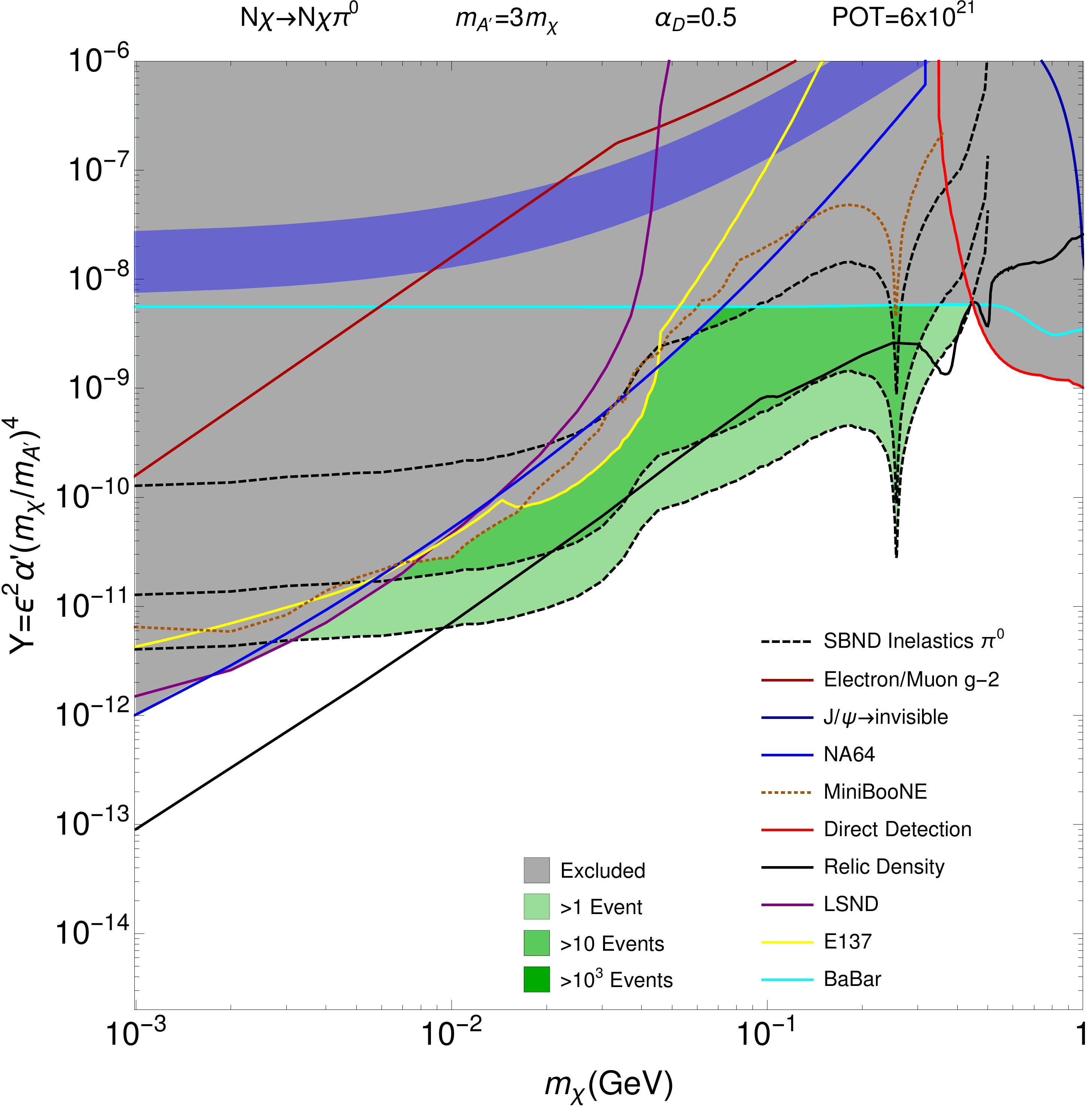}\hspace*{0.3cm}
 \includegraphics[width=0.45\textwidth]{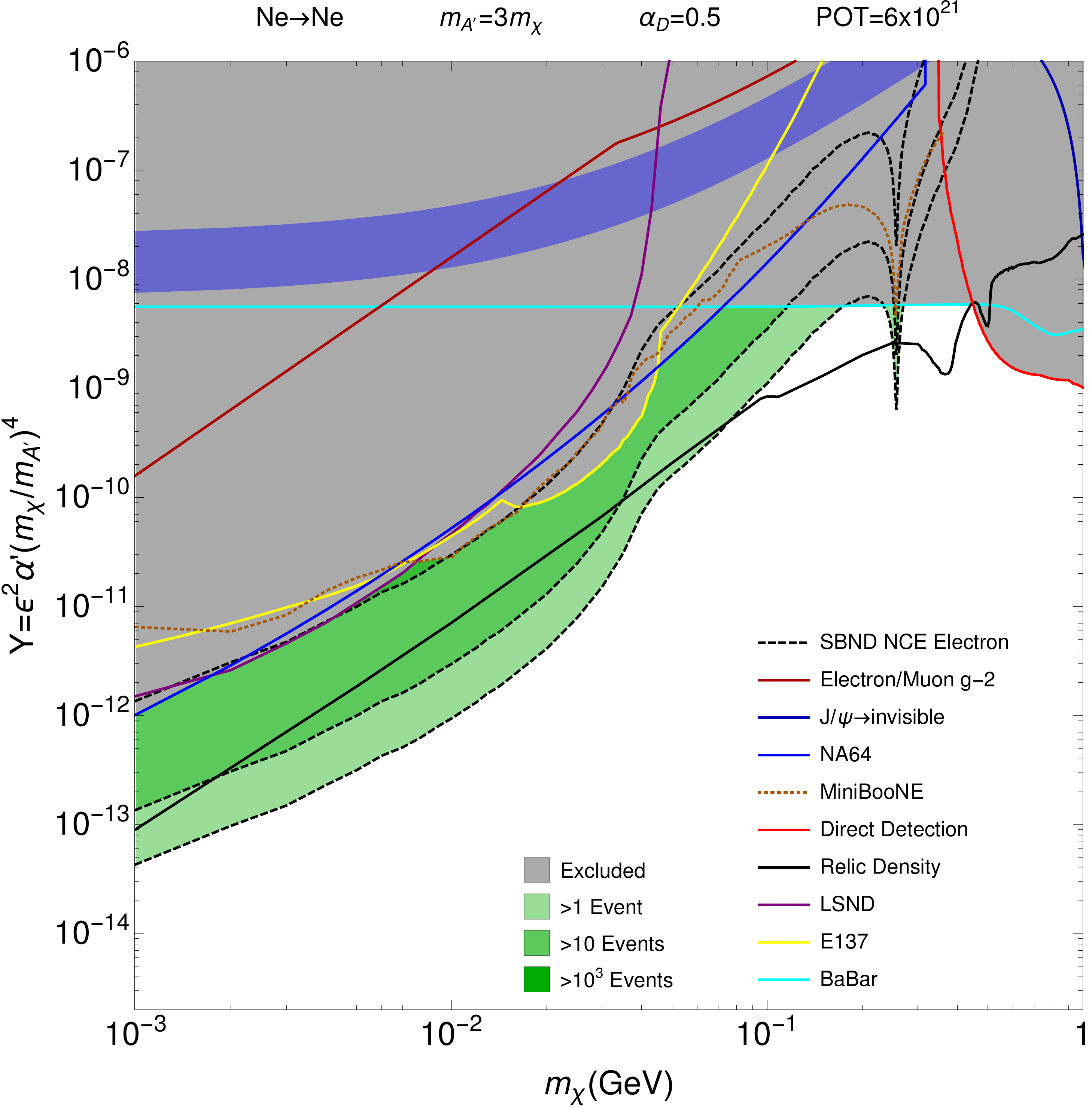}\hspace*{0.3cm}
 }
\caption{\footnotesize Regions of relic abundance parameter (mixing
  strength) $Y$ vs. dark matter mass $m_\chi$ for $6\times10^{21}$ POT that could be
  achieved in a five year run with dedicated proton beam dump medium energy running in the PIP-II era. Left is the signal sensitivity for NC$\pi^0$ and right for NC-electron scattering with the SBND detector at 100 m from the dedicated beam dump.  Both panels show regions where we expect 1--10 (light green), 10--1000 (green), and more than 1000 (dark green) scattering events. The solid black line is the scalar relic density line that can be probed.}
\label{ExpSensitivity1}
\end{figure}

Leveraging the pioneering work of MiniBooNE's dark matter search, it has become clear that a significantly improved sub-GeV dark matter search can be performed with a new dedicated BNB beam dump target station optimized to stop charged pions which produce neutrino backgrounds to a dark matter search.  The new beam dump target can be constructed within 100 m of the SBN Near Detector (SBND) that is currently under construction~\cite{SBN}. In the PIP-II era 8 GeV protons with higher power can be delivered to the BNB, up to  15 Hz and 115 kW, which is a significant increase from the current 5 Hz and 35 kW. In a five year run this would result in $6\times10^{21}$ Proton on Target (POT) delivered to a new dedicated beam dumped while still delivering maximum levels of protons (35 kW) to the neutrino program. The five year sensitivity, shown in Figure~\ref{ExpSensitivity1}, would be greater than an order of magnitude improvement on the MiniBooNE dark matter search due to the reduced neutrino background from the dedicated beam dump, the detector's close proximity to the beam dump, and higher protons on target. Furthermore, new vector and scalar/pseudoscalar dark sector models that couple to leptons or quark currents that explain the MiniBooNE low energy excess make unique prediction that can also be tested with SBND and the BNB beam dump \cite{Dutta:2021cip}.

\section{A Kaon Decay-at-rest Sterile Neutrino Search}

A new BNB beam dump target station also provides a unique opportunity to probe the sterile neutrino oscillation explanation of the MiniBooNE low energy excess using $\nu_\mu$ disappearance. The KPIPE experimental concept, outlined in Ref.~\cite{Axani:2015dha}, calls for a very long (120~m) and thin (1.5~m radius) cylindrical detector close to and oriented radially outward from an intense beam-dump source of monoenergetic 236~Mev $\nu_\mu$ from charged-kaon decay-at-rest ($K^+ \rightarrow \mu^+ \nu_\mu$, with branching ratio of 64\%) to achieve sensitivity to short-baseline $\nu_\mu$ disappearance. The idea is to search for an $L/E$-dependent oscillation wave using fixed-$E$ neutrinos with minimal background and only modest detector requirements. 

The KPIPE detector, relying on liquid scintillator and silicon photomultipliers (or PMTs), is designed to look for 236~MeV $\nu_\mu n \rightarrow \mu^{-} p$ interactions, which provide a unique double-flash coincidence due to the muon decay following the initial prompt event. Mapping these interactions as a function of distance along the detector pipe, with a nominal, no-oscillation expectation of a $1/r^2$ rate dependence, provides sensitivity to muon-flavor disappearance. Given a beam dump, decay-at-rest neutrino source, the beam-based $\nu_\mu$ background (from decay-in-flight mesons) to these signal events is expected to be completely sub-dominant, at the 1-2\% level. While cosmics can be considered a concern for such a surface or near-surface detector, this background can be mitigated by typical accelerator duty factors of $\sim10^{-6}-10^{-5}$ combined with the short charged kaon lifetime (13~ns). The monoenergetic neutrino source, combined with low decay-in-flight background and small beam duty factor, means that the signal-to-background ratio is expected to be well over 50:1 in the scenarios considered. This large ratio means that the detector requirements, in particular the photocoverage, can be quite modest. In fact, a preliminary estimate at Ref.~\cite{cost} predicts that the entire KPIPE detector would cost \$5M.

The KPIPE detector was originally envisioned to be paired with the 3~GeV, 730~kW (currently, with 1~MW planned) J-PARC Spallation Neutron Source. Aside from the primary proton energy, which is above the kaon production threshold, and the high power, this source is particularly attractive because the beam timing structure, two $\sim$80~ns pulses separated by 540~ns at 25~Hz, provides an extremely low duty factor ($4\times 10^{-6}$), essential for cosmic background rejection. The drawback of this source, however, is that the 3~GeV primary proton energy, while above threshold, is somewhat lower than optimal for charged kaon production per unit power: at 3~GeV, the MARS15 software package~\cite{Mokhov:2012nco} predicts 0.007 KDAR $\nu_\mu$/POT. With an increase in proton energy to 8~GeV, for example, the production rate increases by a factor of 10 to 0.07 KDAR $\nu_\mu$/POT. Spatial and facility issues, especially in consideration of the existing materials-science-focused beamlines and experiments, also means that optimal detector placement, with KPIPE calling for a 120~m long detector with closest distance of 32~m from the neutrino source, is challenging.

\begin{figure}[!ht]
\begin{centering}
\includegraphics[width=8.7cm]{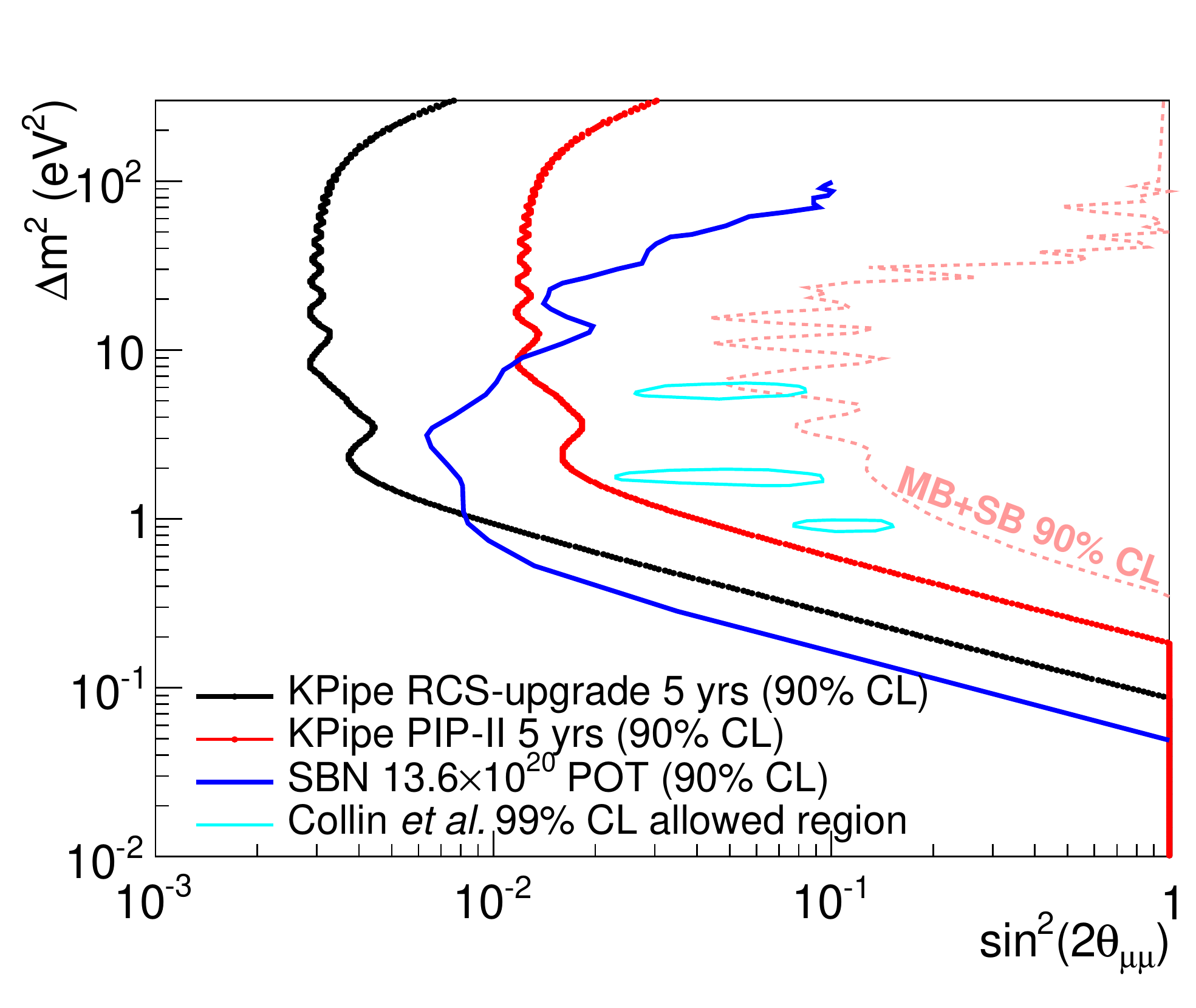}
\vspace{-.6cm}
\caption{The 90\% CL sensitivities of the KPIPE at Fermilab scenarios considered here, in both the PIP-II and RCS upgrade eras. For reference, we also show the expected 90\% CL SBN sensitivity (6~years)~\cite{MicroBooNE:2015bmn}, existing 90\% CL MiniBooNE+SciBooNE limit~\cite{SciBooNE:2011qyf}, and 99\% allowed region from the Collin \textit{et al.} global fit~\cite{Collin:2016aqd}.}
\label{figure_sens}
\end{centering}
\end{figure}

The future Fermilab particle accelerator complex~\cite{Ainsworth:2021ahm}, including PIP-II~\cite{Lebedev:2017vnu} and eventually a new rapid cycling synchrotron (RCS)~\cite{Eldred:2019erg}, can provide an optimal beam-dump/stopped-kaon neutrino source for KPIPE, in terms of beam energy (8~GeV), beam timing ($\sim10^{-5}$ duty factor), and spatial considerations. Using the detector and Fermilab-accelerator assumptions shown in Table~\ref{table:values}, and scaling based on the detailed study in Ref.~\cite{Axani:2015dha}, we expect KPIPE could achieve the sensitivity to short-baseline $\nu_\mu$ disappearance shown in Figure~\ref{figure_sens}. As can be seen, this sensitivity surpasses, and is highly complementary to, SBN (6~years) at $\Delta m^2>10$~eV$^2$ for both scenarios considered and $\Delta m^2>1$~eV$^2$ for the RCS upgrade era case. 

\begin{table} [!ht]
\begin{center}
\begin{tabular}{|c|c|}
\hline
Experimental assumptions &     \\ \hline
Detector length &  120~m   \\
Active detector radius & 1.45~m   \\
Closest distance to source &  32~m   \\
Liquid scintillator density &  0.863 g/cm$^3$  \\
Active detector mass & 684~tons  \\
Primary proton energy & 8~GeV  \\
Target material & Hg or W  \\
KDAR $\nu_\mu$ yield (MARS15)  &  0.07~$\nu_\mu$/POT   \\
$\nu_\mu$ CC $\sigma$ @ 236 MeV (NuWro)  &  $1.3\times10^{-39}~\mathrm{cm}^2/\mathrm{neutron}$   \\
KDAR signal efficiency &    77\% \\
Vertex resolution  &   80~cm \\
Light yield & 4500~photons/MeV \\
Uptime (5 years) & 5000 hours/year \\
$\nu_{\mu}$ creation point uncertainty &    25~cm\\ \hline \hline
PIP-II era assumptions &   \\ \hline
Proton rate (0.08~MW) & 1.0 $\times 10^{21}$ POT/year\\
Beam duty factor & $1.6\times10^{-5}$\\
Cosmic ray background rate &    110~Hz \\ 
Raw KDAR CC event rate &  $2.7\times10^4$ events/year \\ \hline \hline
RCS upgrade era assumptions &   \\ \hline
Proton rate (1.2~MW) & 1.5 $\times 10^{22}$ POT/year\\
Beam duty factor & $5.3\times10^{-5}$\\
Cosmic ray background rate &   360~Hz \\ 
Raw KDAR CC event rate &  $4.0\times10^5$ events/year \\ \hline
\end{tabular} \caption{Summary of the relevant KPIPE experimental parameter assumptions.}\label{table:values}
\end{center}
\end{table}

\section{Required Infrastructure}
The new PIP-II linac will be able to deliver a proton beam of significantly higher power to the Booster than the current linac. This will result in 15 Hz of beam to be delivered to the BNB achieving 115 kW, or about three times the power of current delivery.   The BNB neutrino target and horn have a power limit of 35 kW, which leaves 80 kW of power for other uses. A new target station fed by the BNB, and on axis with the existing SBN neutrino experiment could be built relatively quick and at modest cost.  Such a facility could be run concurrently with the SBN neutrino program, only using protons beyond the 35 kW limit.  Events would be trivially separated on a pulse by pulse basis based on the which target the beam is being delivered too. The facility will require a Fe target about 2 m in length and 1 m in width to absorb the protons and resulting charged pions.  Shielding and cooling requirements up to 80 kW are straightforward.  Such a target would reduce backgrounds by another three orders of magnitude relative the regular neutrino running (see next section for details).  Besides the higher power, the reduced neutrino flux background enables a significantly more sensitive search for dark matter relative to the MiniBooNE beam off target run.

\section{Decay-in-flight Neutrino Flux Reduction with Improved Beam Dump}

\begin{figure}[h]
 \centerline{ 
 \includegraphics[width=0.5\textwidth]{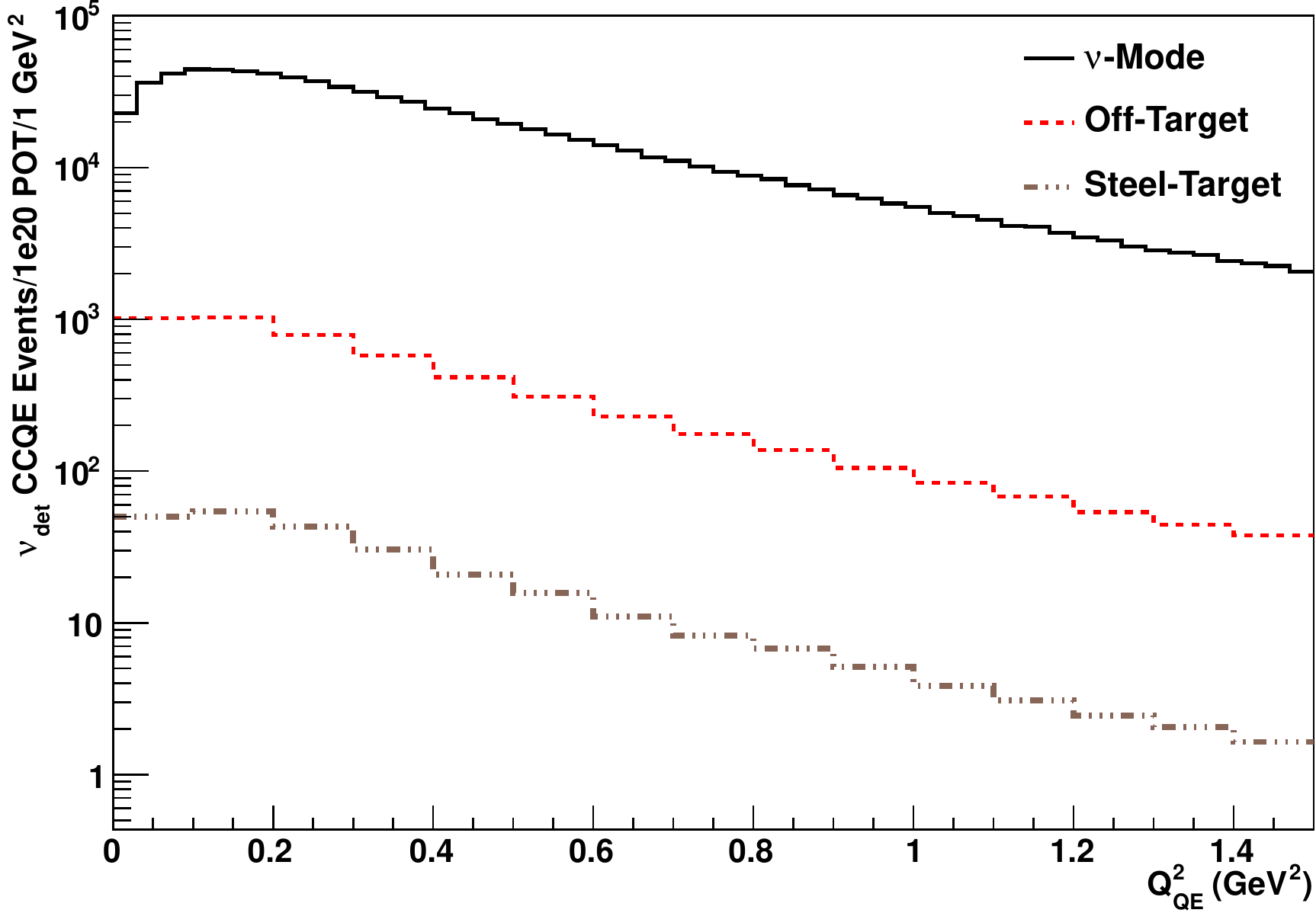}\hspace*{0.3cm}}
\caption{\footnotesize Detailed decay-in-flight neutrino flux estimation for neutrino running 
  (solid black line), beam-off-target 50 m absorber running (dotted red line), 
  and a dedicated new BNB beam dump target station (dotted brown line).   In this final mode, the decay-in-flight neutrino 
  flux reduction is a factor of 1000, or about 20 times better than 
  50 m absorber running. }
\label{NuFlux}
\end{figure}

To leverage the increased dark matter signal rate production, a
corresponding reduction in decay-in-flight neutrino-induced backgrounds is required.
The MiniBooNE-DM beam-off-target run steered the protons past the Be
target/horn and onto the 50 m absorber.  This reduces the
neutrino-induced background rate by a factor of $\sim$50, but there
was still significant production of neutrinos from proton interactions
in the 50 m of decay pipe air and beam halo scraping of the target.
Further reduction of neutrino production occurs by directing the proton beam directly onto a 
dense beam stop absorber made of Fe or W.  This puts the end of
the proton beam pipe directly onto the dump with no air gap.  Detailed BNB dump beam line simulations, which have been verified by data \cite{MB-DM}, demonstrate that this would reduce
neutrino-induced backgrounds by a factor of 1000 over Be-target neutrino
running, which is a factor of twenty better than the 50 m
absorber as demonstrated in Figure \ref{NuFlux}.  

\section{Timescales, Costs, and Similar Facilities}

The timescale for building a BNB beam dump target station is similar to the construction of PIP-II and the expected upgrade in protons once online. The SBN detectors are expected to run for at least 10 years. The new dedicated beam dump could be built sooner and begin running using the SBN detectors at a lower rate until the PIP-II upgrade is complete. Such a facility could be built quickly, 1-2 years, and at modest cost below \$5M.  There are no other similar facilities in the world currently or planned in the next five years that can probe for dark matter masses up to 1 GeV with a proton beam.

\bibliographystyle{unsrt}
\bibliography{bibliography}

\end{document}